\documentclass[twocolumn,superscriptaddress,longbibliography,aps,pra,preprintnumbers]{revtex4-2}

\usepackage[urlcolor=blue,colorlinks=true,citecolor=blue,linkcolor=blue,pdfstartview={FitH},bookmarks=false]{hyperref}
\usepackage[utf8]{inputenc}
\usepackage{graphicx,comment,float}
\usepackage{xcolor,soul}
\usepackage{dcolumn}
\usepackage{bm}
\usepackage{amsmath}
\usepackage{amssymb}
\usepackage{mathrsfs} 
\usepackage[normalem]{ulem}
\usepackage{url}

\usepackage{xcolor}
\definecolor{mygreen}{rgb}{0.0, 0.6, 0.0}
\definecolor{pjorange}{rgb}{0.8, 0.3, 0.0}
\definecolor{jlblue}{rgb}{0.2, 0.5, 0.7}

\begin{document}


\title{Superexchange Interaction in Insulating EuZn$_{2}$P$_{2}$}

\author{Karan Singh}
\affiliation{Institute of Low Temperature and Structure Research, Polish Academy of Sciences, Okólna 2, 50-422 Wrocław, Poland}

\author{Shovan Dan}
\email[e-mail: ]{dan.shovan@gmail.com}
\affiliation{Institute of Low Temperature and Structure Research, Polish Academy of Sciences, Okólna 2, 50-422 Wrocław, Poland}

\author{A. Ptok}
\affiliation{Institute of Nuclear Physics, Polish Academy of Sciences, W. E. Radzikowskiego 152, PL-31342 Kraków, Poland}

\author{T. A. Zaleski}
\affiliation{Institute of Low Temperature and Structure Research, Polish Academy of Sciences, Okólna 2, 50-422 Wrocław, Poland}

\author{O. Pavlosiuk}
\affiliation{Institute of Low Temperature and Structure Research, Polish Academy of Sciences, Okólna 2, 50-422 Wrocław, Poland}

\author{P. Wiśniewski}
\affiliation{Institute of Low Temperature and Structure Research, Polish Academy of Sciences, Okólna 2, 50-422 Wrocław, Poland}

\author{D. Kaczorowski}
\email[e-mail: ]{d.kaczorowski@intibs.pl}
\affiliation{Institute of Low Temperature and Structure Research, Polish Academy of Sciences, Okólna 2, 50-422 Wrocław, Poland}

\date{\today}

\begin{abstract}
We report magnetic and transport properties of single-crystalline EuZn$_{2}$P$_{2}$, which has trigonal CaAl$_2$Si$_2$-type  crystal structure and orders antiferromagnetically at $\approx$23~K. 
Easy $ab$-plane magneto-crystalline anisotropy was confirmed from the magnetization isotherms, measured with a magnetic field applied along different crystallographic directions ($ab$-plane and $c$-axis). Positive Curie-Weiss temperature indicates dominating ferromagnetic correlations.  
Electrical resistivity displays insulating behavior with a band-gap of $\approx\,$0.177~eV, which decreases to $\approx\,$0.13~eV upon application of a high magnetic field.
We explained the intriguing presence of magnetic interactions in an intermetallic insulator by the mechanism of extended superexchange, with phosphorus as an anion mediator, which is further supported by our analysis of the charge and spin density distributions. We constructed the effective Heisenberg model, with exchange parameters derived from the \textit{ab initio} DFT calculations, and employed it in Monte-Carlo simulations, which correctly reproduced the experimental value of N\'eel temperature.
\end{abstract}

\maketitle

\section{Introduction}
Interplay between magnetism and band topology generates alluring magnetic and electronic properties, such as superconductivity~\cite{fassler1997basn3}, non-trivial electronic states, and anomalous Hall effect~\cite{Xu.19, Yan.22} in Zintl-phase materials. 
Most interestingly, magnetically coupled topological charge carriers can be modulated by changing the spin structure with an application of external magnetic field. 
Eu-based Zintl-phase materials are in focus for exhibiting quantum phenomena as well as non-trivial topological properties, with potential applications in thermoelectric and spintronic devices~\cite{Xu.19,Yan.22,chen2019zintl,zada2021structure}. 
Eu$T_2X_2$ ($T=$ transition metal and $X=$ pnictogen) compounds have been argued to have various magnetic ground states.
For example, a collinear antiferromagnetic (AFM) structure has been reported for EuCd$_2$As$_2$, EuZn$_2$As$_2$, and EuMg$_2$Bi$_2$~\cite{rahn.soh.18, Wang2022,Pakhira2020}, whereas a helical AFM structure has been predicted for EuIn$_2$As$_2$~\cite{Xu.19}, which possesses robust non-trivial topological features. 
Field evolution of the magnetic structure has been reported for EuCd$_2$As$_2$ and EuMg$_2$Bi$_2$ - diffraction studies (resonant X-ray and neutron, respectively) have shown that below the N\'{e}el temperature, $T_{\rm N}$, the Eu$^{2+}$ moments are ferromagnetically aligned within the $ab$-plane and antiferromagnetically coupled with those in the adjacent $ab$-plane, forming an A-type AFM structure~\cite{rahn.soh.18,Pakhira2021}. 
Applied magnetic field breaks the collinearity of AFM structure, turns spin arrangement into noncollinear one, and finally in polarized paramagnetic state~\cite{rahn.soh.18,Wang2022,Soh2020}. It is worth to mention here, that EuMg$_2$Bi$_2$, EuCd$_2$As$_2$, EuZn$_2$As$_2$, EuZn$_2$Sb$_2$, EuCd$_2$Sb$_2$, and EuZn$_2$P$_2$ all have the same trigonal crystal structure (space group $P\bar{3}m1$)~\mbox{\cite{Wang2022,rahn.soh.18,weber.06,Soh2020,berry.stewart.22}}. 
Notwithstanding, large variations in the magnetoconductivity were observed, manifesting Dirac state in EuCd$_2$As$_2$~\cite{rahn.soh.18,Soh2020}, semiconducting/insulating state in EuZn$_2$P$_2$~\cite{berry.stewart.22} or a flat band in EuZn$_2$As$_2$~\cite{Wang2022}.

Coexistence of insulating behavior and the AFM order makes EuZn$_2$P$_2$ a very interesting system.
Magnetic and electronic transport properties of that compound have recently been reported, and the authors claimed that the compound is an A-type antiferromagnet with $T_{\rm N}$ ($\approx~23$ K) and only $\frac{2}{3}$ of the magnetic entropy expected for Eu$^{2+}$ ion ($\mathcal{R}\ln8$) recovered above $T_{\rm N}$~\cite{berry.stewart.22}. 
An insulating behavior of the compound (with a band gap of $0.11$\,eV) was observed from electrical transport as well as DFT calculations. 
In the same article, it has been reported that the magnetic easy-axis is the crystallographic $c-$axis~\cite{berry.stewart.22}. 
In contrast, most of the materials that crystallize in a structure similar to EuZn$_{2}$P$_{2}$ have the easy axis in the $ab$-plane and the hard axis along the $c-$axis~\cite{Wang2022,rahn.soh.18,weber.06,Soh2020}.

In this context, we revisited the magnetization, electrical resistivity and specific heat of high-quality single crystal to verify the easy- and hard-axes and the reason for unrecovered magnetic entropy. Notably, we established that the underlying magnetic interaction is governed by the exchange mechanism but not arising from the dipolar interactions as proposed in the Ref.~\cite{berry.stewart.22}. To support our argument, we performed an \textit{ab initio} DFT calculation, and extracted the anisotropic exchange parameters. Using the exchange parameters, we simulated the specific heat and magnetization as a function of temperature using Monte Carlo method with the effective Heisenberg model.

Primary results presented in this paper are organized in the following manner. 
A detailed description of the used methods can be found in Sec.~\ref{sec.methods}. The details of the crystal structure and electronic band structure obtained from the DFT calculations are presented in Sec.~\ref{sec.crystal}.
In Sec.~\ref{ssec:mag}, we have characterized the magnetic properties of the compound along the principal crystallographic directions.  
The specific heat study and electrical resistivity are described in Sec.~\ref{ssec:rho} and Sec.~\ref{ssec:cp}, respectively. 
The Monte Carlo (MC) simulation of the magnetic ordering and specific heat using the exchange parameters obtained from DFT calculations, are shown in Sec.~\ref{ssec:theo}. 
To understand more the magnetic interactions among Eu atoms, a comprehensive magnetocaloric effect study is shown in the Supplemental Material (SM)~\cite{SuppMat}.
Finally, we conclude our findings in Sec.~\ref{sec.sum}.

 \begin{figure*}[!]
    \centering
    \includegraphics[width=0.9\textwidth]{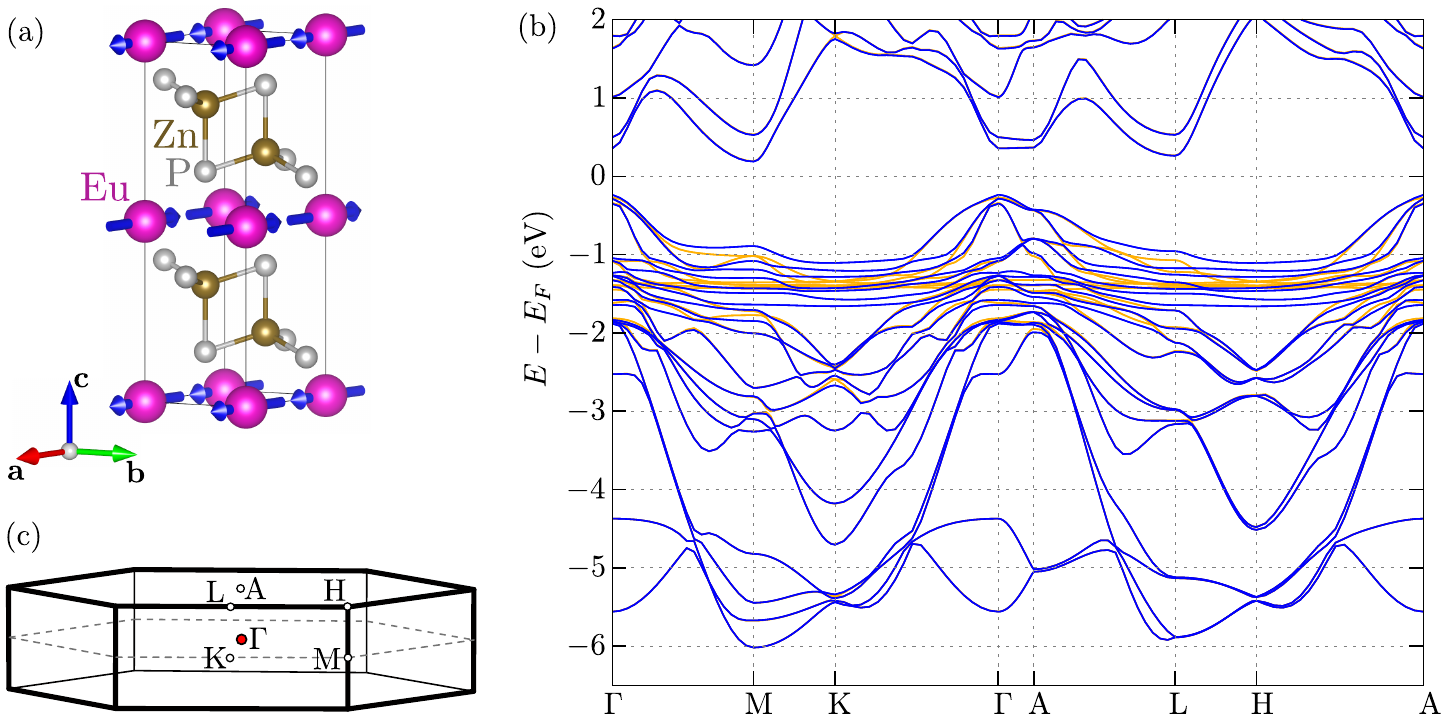}
    \caption{The crystal structure of EuZn$_{2}$P$_{2}$ with $P\bar{3}m1$ symmetry and A-type AFM order with in-plane magnetic moments (a). 
    Electronic band structure for Eu $4f$ electrons treated as valence states (b) along high symmetry direction of the Brillouin zone (c).
    The orange and blue lines present results in the absence and presence of the spin--orbit coupling, respectively.}
    \label{fig:band}
\end{figure*}
 \section{Methods}
\label{sec.methods}

\subsection{Experimental}

Single crystals of EuZn$_{2}$P$_{2}$ were grown from Sn flux. Starting materials including Eu, Zn, red P, and Sn were mixed in a molar ratio 1:2:2:45.
The mixture was sealed in a quartz ampule and heated up very slowly (about 5 days) to 1100$^{\circ}$C, kept at that temperature for 24 hours and then cooled to 800$^{\circ}$C (at 2$^{\circ}$C/h rate), when the flux was removed by centrifuging. 

Composition of obtained single crystals was examined with a scanning electron microscope (SEM) (FEI Technologies) with an energy dispersive X-ray spectrometer (EDS) (Genesis XM4). Single crystals' quality and their orientation were identified using the Laue back-scattering diffraction (Laue-COS, Proto Manufacturing). We took special care in the orientation of the crystal and precisely assigned the crystallographic directions.
Magnetization measurements were carried out on the oriented single crystals using a SQUID magnetometer (MPMS-XL, Quantum Design). 
The electrical resistivity measurements were performed on a PPMS platform (Quantum Design) by a conventional 4-probe method. A rectangular-prism-shaped sample ($1.36 \times 1\times 0.8$ mm$^3$) was cut from an oriented single-crystal and electrical contacts were made with $50$~$\mu$m gold wire and silver paint. Measurements were performed in the magnetic field applied parallel to the $c$-axis, while an electric current of $3$~mA was  within the $ab$-plane. We restricted our electrical transport measurements to temperatures above 120 K, due to the instrumental limitations of measuring high resistance.

\subsection{Computational details}
\label{sec.dft}

The first-principles DFT calculations were performed using the projector augmented-wave (PAW) potentials~\cite{bloch.94} implemented in the {\it Vienna Ab initio Simulation Package} ({\sc Vasp}) code~\cite{kresse.hafner.94, kresse.furthmuller.96, kresse.joubert.99}.
The calculations were performed within the generalized gradient approximation (GGA) in the Perdew, Burke, and Ernzerhof (PBE) parameterization~\cite{pardew.burke.96}.

In this study, we used the experimental lattice parameters
while the atoms positions were optimized. 
The magnetic unit cell containing two formula units was optimized, with the $12 \times 12 \times 4$ {\bf k}--point $\Gamma$-centered grid in the Monkhorst--Pack scheme~\cite{monkhorst.pack.76}.
For converging  the  optimization loop, we took the energy difference 
of $10^{-6}$~eV and $10^{-8}$~eV for ionic and electronic degrees of freedom, respectively.
The calculations were performed with the energy cut-off set to $400$~eV.
In calculations, the Eu $4f$ electrons were treated as valence states.
Correlation effects were introduced within DFT+U, proposed by Dudarev {\it et al.}~\cite{dudarev.botton.98}.
Similar to  previous study~\cite{berry.stewart.22}, we assumed $U = 5 $~eV as a realistic value for the Eu $4f$ states.
\section{Results and discussions}

\subsection{Crystal structure and electronic band structure}
\label{sec.crystal}

EuZn$_{2}$P$_{2}$ crystallizes in CaAl$_{2}$Si$_{2}$-type structure ($P\bar{3}m1$ space group, No. 164), with $a = 4.087$~\AA, and $c = 7.010$~\AA~\cite{frik.mewis.99}.
The atoms are located at the high symmetry Wyckoff positions: Eu $1a$ (0,0,0), Zn $2d$ (2/3,1/3,$z_\text{Zn}$), and P $2d$ (1/3,2/3,$z_\text{P}$).
After optimization, we found free parameters as $z_\text{Zn} = 0.3683$ and $z_\text{P} = 0.2704$, which are close to reported in previous study ($z_\text{Zn} = 0.3695$ and $z_\text{P} = 0.2692$)~\cite{berry.stewart.22}.

Similarly to EuCd$_{2}$As$_{2}$~\cite{rahn.soh.18}, at low temperatures the system realizes A-type AFM structure with moments in $ab$-planes [Fig.~\ref{fig:band}(a)].
From the DFT calculations, the Eu magnetic moment is nearly $7$~$\mu_{\rm B}$ (intrepidity by used $U$ value), which corresponds to $J = S = 7/2$ of Eu$^{2+}$ ion. 
The electronic band structure, shown in Fig.~\ref{fig:band}(b), is characterized by the indirect gap, realized between $\Gamma$ and M points. 
Similar behavior was earlier reported for EuMg$_{2}$Sb$_{2}$~\cite{Santanu22}.
Values of the indirect and direct gaps strongly depend on the treatment of $4f$ electrons, as well as a correlation parameter $U$ (see Tab.~S1 in the SM~\cite{SuppMat}), due to the band structure renormalization, similar to that reported for the rare-earth nickelates~\cite{ptok.basak.23}.
The band gap reported experimentally is around $0.1$~eV (cf. Electrical transport and \cite{berry.stewart.22}), consistent with DFT calculations. 

\paragraph*{Role of $4f$ electrons and correlation.}
We compare the electronic band structure for Eu $4f$ electrons treated as a core state and valence state. 
In both cases, the electronic band structure exhibits similar features (cf.~Fig.~S3 and Fig.~S4 in the SM~\cite{SuppMat}).
However, treating $4f$ electrons as a core states led to the strong overestimation of band gaps (cf. Tab.~S1 in the SM~\cite{SuppMat}).
Including the Hubbard-type local interaction $U$ in the DFT calculations (DFT+U) shifted the $4f$ valence state to lower energies (see Fig.~S4 in the SM~\cite{SuppMat}).
We wish to mention here, that the Eu $4f$ states for assumed $U = 5$~eV are located around the binding energy $1.5$~eV, and are well visible as the flat bands [Fig.~\ref{fig:band}(b)].
Additionally, from the comparison of band structures calculated without and with spin--orbit coupling, we can conclude that this  coupling strongly affects only the Eu $4f$ states.
This is reflected in the strong decoupling of the flat band around binding energy $1.5$~eV (cf. orange and blue lines on Fig.~\ref{fig:band}(b), related to the results in the absence and presence of the spin--orbit coupling, respectively).

\begin{figure*}
    \centering
    \includegraphics[width=\textwidth]{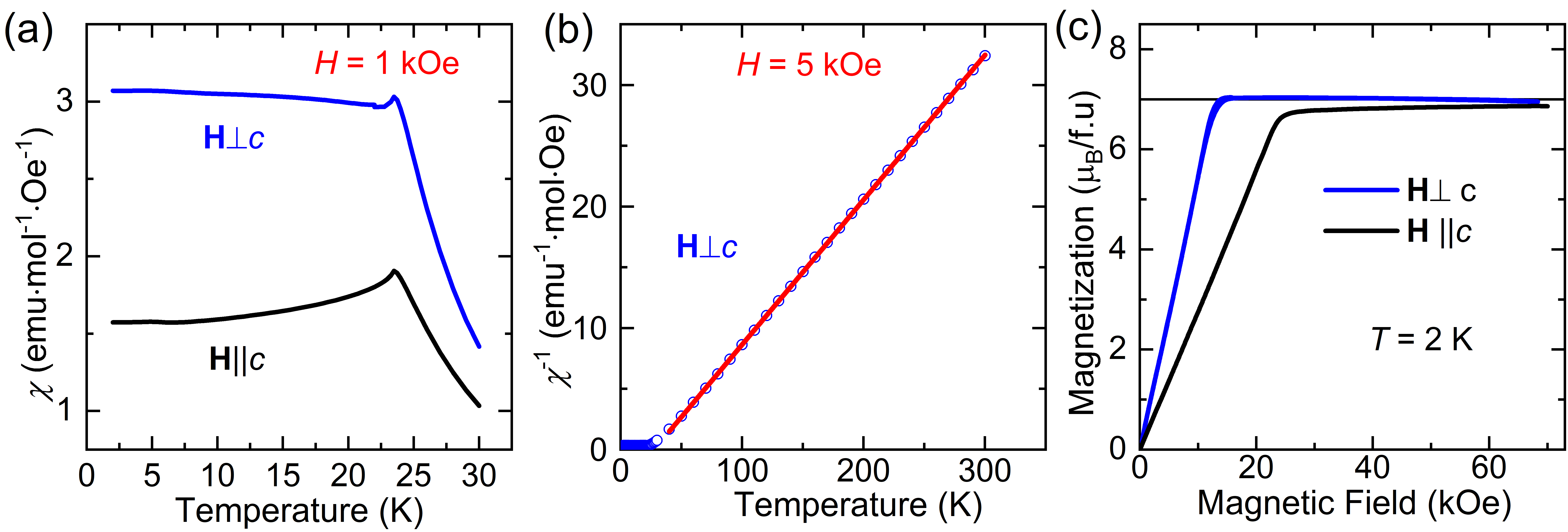}
    \caption{(a) Magnetic susceptibility measured in magnetic field of $1$~kOe, applied parallel and perpendicular to the $c$-axis. (b) Inverse susceptibility vs. T, (in $H = 5\,\rm kOe$), red line represents fit with the Eq.~(\ref{eqn:CW}). (c) Magnetization isotherms, measured at $T = 2$~K in magnetic field applied parallel and perpendicular to the $c$-axis. The horizontal line corresponds to the $7\mu_{\rm B}$ of Eu$^{2+}$ ion.}
    \label{fig:MT}
\end{figure*}

\subsection{Magnetization}
\label{ssec:mag}

Fig.~\ref{fig:MT}(a) shows the variation of static magnetic susceptibility ($\chi= M/H$) with temperature (2--30~K) in a magnetic field of $1$~kOe applied parallel and perpendicular to the crystallographic $c$-direction. The compound orders antiferromagnetically below $T_{\rm N}=$23.5~K, which is consistent with that reported earlier~\cite{berry.stewart.22}.  
The $\chi(T)$ measured in $5$~kOe (\textbf{H}$\perp$\textit{c}) is fitted with the Curie--Weiss law, in the temperature range 30--300\,K:
\begin{equation}
    \chi = \frac{\rm C}{T-\theta_{p}} ,
    \label{eqn:CW}
\end{equation}
\noindent where C is the Curie constant and $\theta_{p}$ is the paramagnetic Curie--Weiss temperature [as shown with red line in Fig.~\ref{fig:MT}(b)]. 
The fit of $\chi^{-1}(T)$ with Eq.~(\ref{eqn:CW}) provided an effective magnetic moment, $\mu_{\rm eff}=8.19\,\mu_{\rm B}$, and $\theta_{p}=$27.45~K. 
The value of $\mu_\mathrm{eff}$ is close to the theoretical effective magnetic moment of Eu$^{2+}$ ion [$g_J\sqrt{J(J + 1)} = 7.94$~$\mu_{\rm B}$]. 
The large positive value of $\theta_{p}$ suggests the dominating presence of ferromagnetic (FM) interactions.

Fig.~\ref{fig:MT}(c) shows the magnetic field dependent magnetization $M(H)$ measured at $2$~K for both \textbf{H}$\perp$\textit{c} and \textbf{H}$\parallel$\textit{c} directions. For \textbf{H}$\parallel$\textit{c}, magnetization increases continuously with increasing magnetic field up to $H_{sat}^{\parallel} \approx$24~kOe, where a saturation is observed. The saturated moment is 6.8~$\mu_{\rm B}$, which is very close to the theoretical value for spin-only magnetic moment of Eu$^{2+}$ ion (i.e. 7$~\mu_{\rm B}$).
For \textbf{H}$\perp$\textit{c} the saturated moment is the same, although the saturating field is significantly smaller: $H_{sat}^{\perp}\approx 14$~kOe. 

To characterize the underlying magnetic interactions in the EuZn$_{2}$P$_{2}$, $\chi(T)$ in different magnetic fields were collected as shown in Fig.~S6(a, b) in the SM~\cite{SuppMat}. In both \textbf{H}$\parallel$\textit{c} and \textbf{H}$\perp$\textit{c}, $T_{\rm N}$ shifts towards lower temperatures and transition gradually broadens with the increasing magnetic field. Fig.~S7(a, b) in the SM~\cite{SuppMat}, shows the magnetic isotherms, $M(H)$ measured in the temperature range $5$--$30$~K in the field of both orientations.  A gradual change from the AFM-typical saturation to the Brillouin-type behavior is observed while increasing the temperature in the ordered region.  

\begin{figure}
    \centering
    \includegraphics[width=0.9\columnwidth]{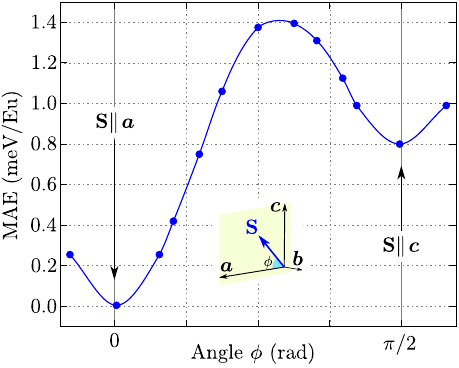}
    \caption{Magneto-crystalline anisotropy energy (MAE) of Eu ions obtained with DFT calculations.
    The orientation of the magnetic moment is denoted by the angle relative to the $a$ direction within the $ac$ plane (inset).
    Minimum at $\phi = 0$ corresponds to the magnetic moments parallel to $a$ direction.}
    \label{fig:mae}
\end{figure}

Next, we calculated magneto-crystalline anisotropy energy (MAE), presented in Fig.~\ref{fig:mae}.
In the calculations, we assumed that the Eu magnetic moment with A-type AFM order is locked in the $ac$ plane, while angle $\phi$ is relative to $a$ direction.
Fig.~\ref{fig:mae} shows the minimum of the MAE when magnetic moments are parallel to the $a$ direction ($\phi = 0$). 
The state with magnetic moments parallel to the $c$ direction ($\phi = \pi/2$) has larger energy.
These two configurations are separated by the energy difference $\approx 0.8$~meV per Eu atom.
Our results strongly suggest that the $ab$ is the easy plane, which differs from the previous report~\cite{berry.stewart.22}.
This finding is in agreement with very recent study of EuZn$_{2}$P$_{2}$, which also claims that the hard axis is parallel to $c$ crystallographic direction~\cite{krebber.kopp.23}.

\subsection{Specific heat study}
\label{ssec:cp}
\begin{figure}[!h]
    \centering
    \includegraphics[width=0.9\columnwidth]{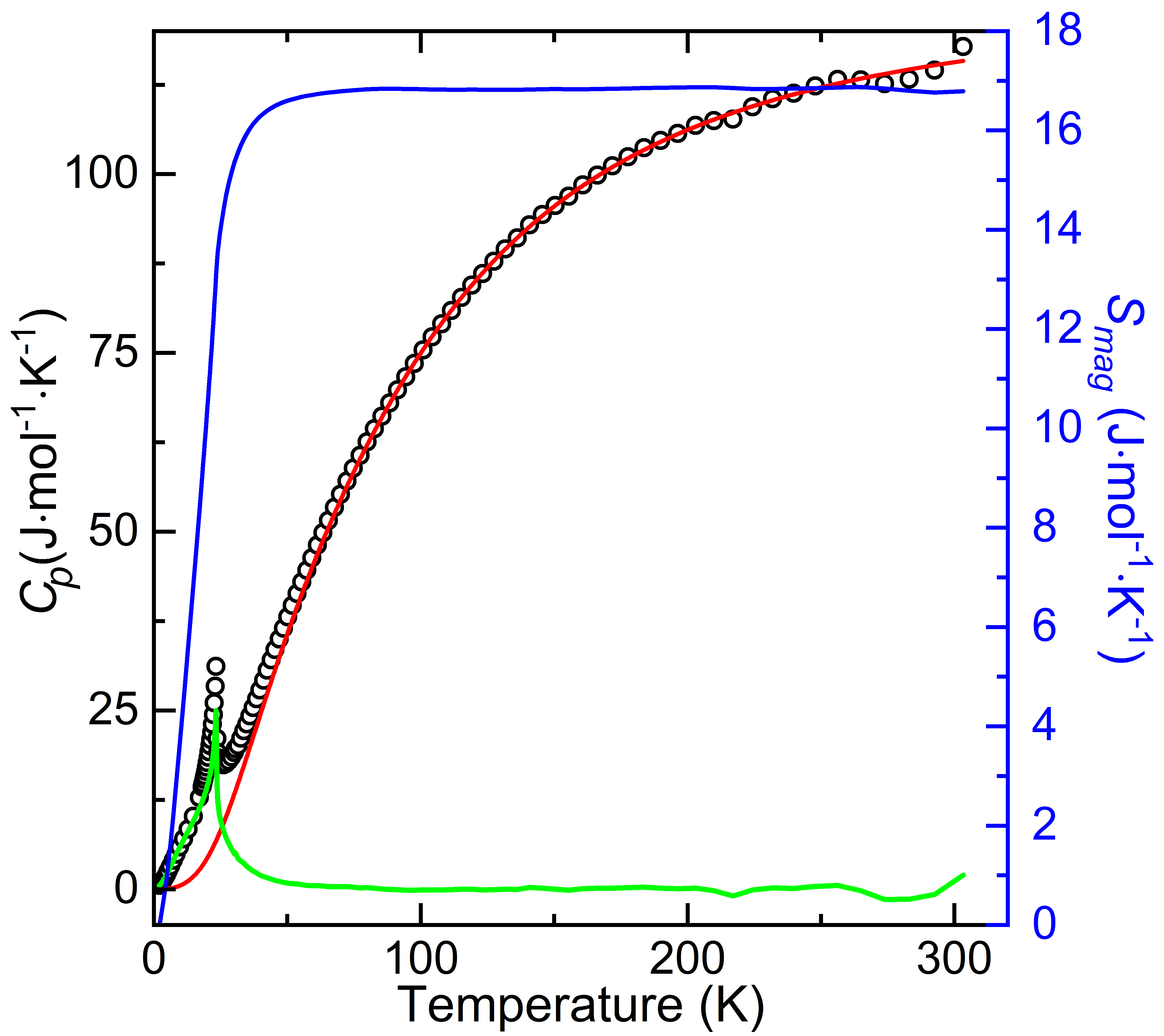}
    \caption{Temperature dependence of the specific heat of EuZn$_{2}$P$_{2}$ (left axis). The solid red curve is the Debye-Einstein fit with Eq.~(\ref{eqn:cp}). The green curve represents the residuals of that fit. The blue line (right axis) shows the corresponding magnetic entropy.}
    \label{fig:cp}
\end{figure}
\begin{figure*}
    \centering
  \includegraphics[width=0.7\textwidth]{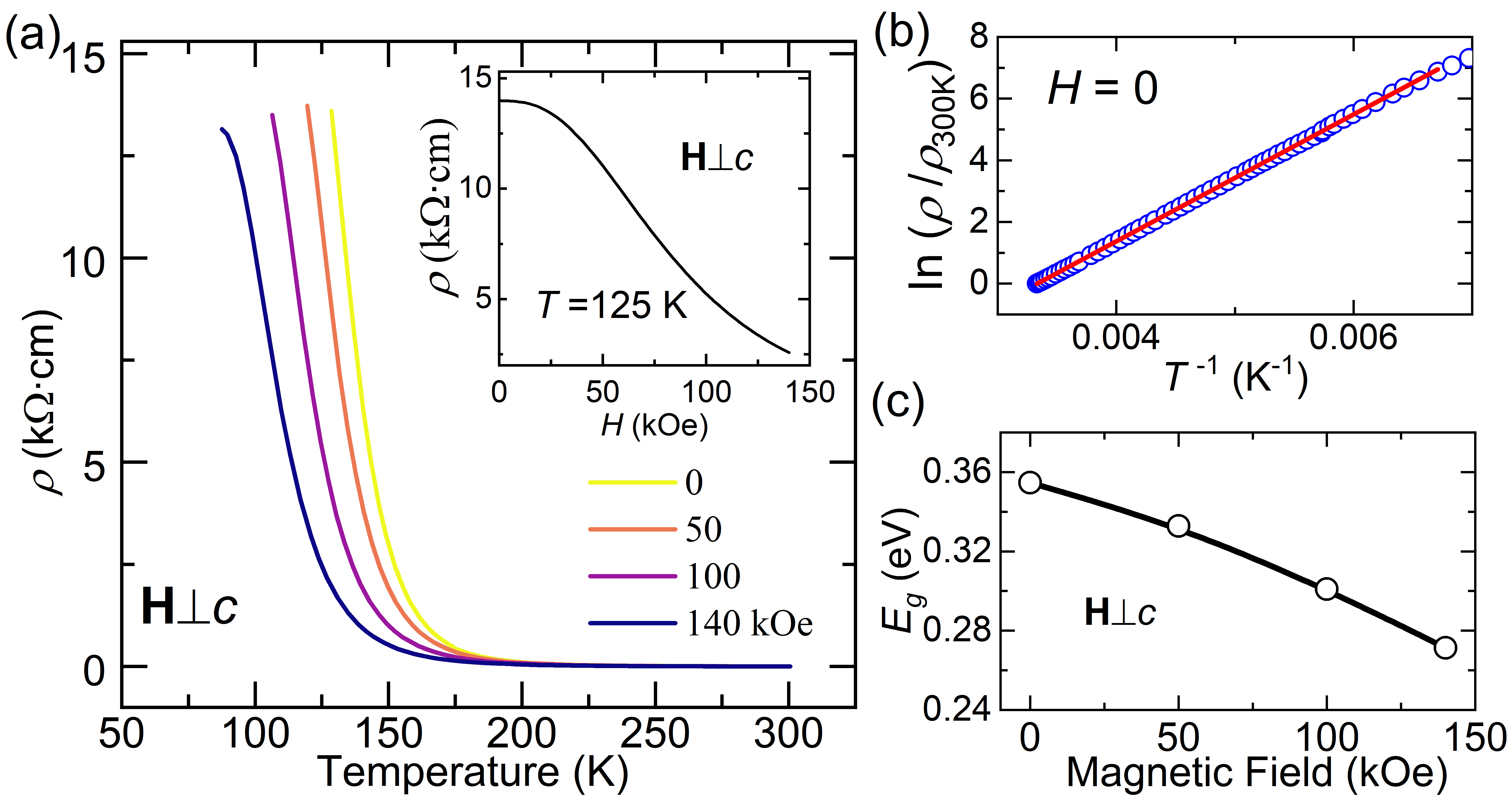}
    \caption{Resistivity vs. temperature in different magnetic fields transverse to the crystallographic $c$-axis and with current flowing within $ab$-plane. Inset: Magnetic field dependence of the resistivity at $125$\,K. (b) $\ln \rho$ vs. $1/T$ plot, for zero applied field, and 150--300~K range of $T$. The red curve represents the fit with Eq.~(\ref{eqn:bandgap}). (c) Band gap as a function of the magnetic field.
    }
    \label{fig:rho}
\end{figure*}
Fig.~\ref{fig:cp} shows the temperature dependence of heat capacity $(C_p)$ measured in zero magnetic fields, which reveals a sharp transition close to $T_{\rm N}$. Above 50\,K, $C_p$ is fitted with a combination of Debye and Einstein terms, and the result is used to extract the magnetic entropy of the compound in the entire temperature range. The Debye model comprises acoustic phonons, whereas the Einstein model tackles optical ones. The model we adopted\cite{gopal2012specific}:
\begin{equation}
 \begin{split}
    C_{ph}(T) = (1-m)\,9\mathcal{R}n\left(\frac{T}{\theta_{\rm D}}\right)^3\int_{0}^{\theta_{\rm D}/T}\frac{x^4e^x{\rm d}x}{{{(e}^x-1)}^2}\\
     + m\,3\mathcal{R}n \left(\frac{\theta_{\rm E}}{T}\right)^2 \frac{e^{\theta_{\rm E}/T}}{\left(e^{\theta_{\rm E}/T}-1\right)^2},
     \label{eqn:cp}
\end{split} 
\end{equation}
where $m$ is the relative weight of the Einstein term, $n$, $\mathcal{R}$, $\theta_{\rm D}$ and $\theta_{\rm E}$ are: the number of atoms per formula unit, universal gas constant, Debye temperature, and Einstein temperature, respectively. Fit of the Eq.~(\ref{eqn:cp}) provided the values of $m$, $\theta_{\rm D}$ and $\theta_{\rm E}$: 0.44(3), $479.88(2)$~K and $135.54(1)$~K, respectively.
$\theta_{\rm D}>\theta_{\rm E}$ indicates the dominance of acoustic phonons at higher temperatures. 
The magnetic entropy ($S_{mag}$) was calculated as $S_{mag} = \int_{2 \rm K}^{300 \rm K}{\frac{1}{T}}(C_p-C_{ph}) {\rm d}T.$ It increases with increasing $T$ above 2\,K, and reaches a maximum of 17~Jmol$^{-1}$K$^{-1}$ above $T_{\rm N}$ (see Fig.~\ref{fig:cp}), which is very close to theoretical entropy $\mathcal{R}\ln(2S+1) =$ 17.3~Jmol$^{-1}$K$^{-1}$ for Eu$^{2+}$ ($S$ = 7/2). 
The earlier report on the same compound had shown that the saturation magnetic entropy attained a maximum of $\frac{2}{3}\mathcal{R}\ln8$, which is much smaller than expected for the Eu$^{2+}$~\cite{berry.stewart.22}. 
The analysis in that paper has employed two-terms Debye model without including the Einstein model. We fitted the same model to our specific heat data (see Fig.~S9 in the SM~\cite{SuppMat}), but it did not retrieve the full magnetic entropy. In contrast, our fit of Eq.~\ref{eqn:cp} (including both the Debye and Einstein model) shows that the saturation entropy is very close to the $\mathcal{R}\ln 8$.  

Above $T_{\rm N}$ there are short-range spin fluctuations present, and hence we restricted the fitting range to $50$--$300$~K. The same model of specific heat was adapted for several Eu$T_2X_2$-type compounds~\cite{Pakhira2020,Santanu22}.

\subsection{Electrical transport}
\label{ssec:rho}
The temperature dependence of the electrical resistivity ($\rho$) measured with the electric current flowing in the hexagonal $ab$-plane ({\bf j}$\perp$\textit{c}-axis) at different magnetic fields is shown in Fig.~\ref{fig:rho}(a). At 300~K, the $\rho$ is $3.35$~$\Omega$cm, with decreasing temperature it increases, and below 128~K the compound becomes an insulator. In magnetic fields up to 140~kOe, the curves appear similar and the magnitude of $\rho$ decreases, but the system remains insulating. The inset of Fig.~\ref{fig:rho}(a) shows field-dependent resistivity $\rho(H)$, at 125~K, which tends to decrease due to reducing band gap. To estimate the band gap ($E_g$) the Arrhenius model was applied by fitting the resistivity data with an equation:
\begin{equation}
  \rho \propto e^{E_g/2k_{\rm B}T} .
  \label{eqn:bandgap}
\end{equation}
Dependence of the $\ln\rho$ on the inverse temperature is shown in Fig.~\ref{fig:rho}(b), which is linear, and yielded $E_g=0.177$~eV when fitted with Eq.~(\ref{eqn:bandgap}).
The band gap decreases monotonically with the application of a magnetic field [Fig.~\ref{fig:rho}(c)] down to $0.13$~eV in $140$~kOe. 
The small difference between the experimentally observed and the theoretically predicted band gap may be due to the shift of Fermi level induced by the defect level present in the charge gap~\cite{berry.stewart.22}.

\begin{figure*}
    \centering
    \includegraphics[width=\textwidth]{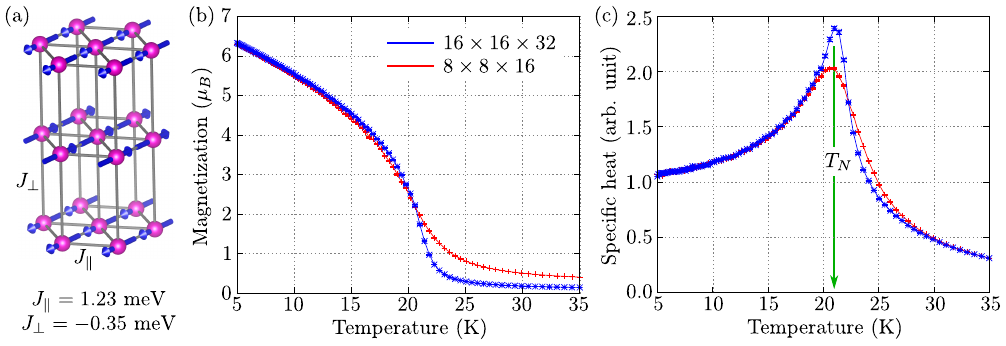}
    \caption{
   (a) Coupling exchange parameters found from {\it ab initio} calculations.
    (b) The average magnetization of a single Eu layer and (c) specific heat obtained from the Monte Carlo simulations. 
    Red and blue lines, correspond to the results obtained for $8 \times 8 \times 16$ and $16 \times 16 \times 32$ spin blocks, respectively.
}
\label{fig:MC}
\end{figure*}

\begin{figure}[!b]
    \centering
    \includegraphics[width=\columnwidth]{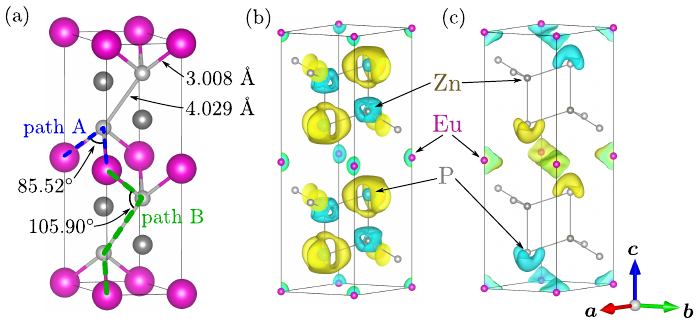}
    \caption{
    (a) Bond lengths and bond angles in EuZn$_{2}$P$_{2}$.
    Blue and green dashed lines show paths participating in the extended superexchange mechanism.
    Modification of the (b) charge and (c) spin  density distribution in EuZn$_{2}$P$_{2}$ obtained from DFT calculations.
    Yellow and cyan isosurfaces indicate the accumulation and depletion of the charge/spin density, respectively.}    
    \label{fig:dens}
\end{figure}

\subsection{Effective Heisenberg model and mechanism of magnetic interaction}
\label{ssec:theo}

To study the magnetic ordering of EuZn$_{2}$P$_{2}$ theoretically, we formulated the effective Heisenberg model~\cite{pajda.kudrnosky.01},
\begin{eqnarray}
\mathcal{H}_{\rm eff} = - \sum_{<i,j>} J_{ij} {\bm e}_{i} {\bm e}_{j} ,
\end{eqnarray}
where ${\bm e}_{i}$ is the unit vector in the direction of the magnetic moment at site $i$, 
and $J_{ij}$ are the direction-dependent exchange parameters.
In this notation, the value of on-site atomic magnetic moment $S_{i}$ is included into the exchange parameters $J_{ij}$ and the summation runs only once over each pair of $i$ and $j$. 

The model concerns only the magnetic moments of Eu, which form two-dimensional triangular layers stacked over each other without any interlayer offset [see Fig.~\ref{fig:MC}(a)].
From the DFT calculations, we found that the triangular layers possess the ferromagnetic coupling $J_{\parallel} = 1.23$~meV, while the AFM interlayer coupling $J_{\perp} = -0.35$~meV (detailed procedure can be found in the SM~\cite{SuppMat}).
The obtained values of the exchange parameters are comparable with those reported in the literature~\cite{guo.wang.20}.
The magnitude of ferromagnetic (intralayer) coupling is bigger than the AFM (interlayer) coupling ($|J_{\parallel}| > |J_{\perp}|$).
This relation between exchange coupling parameters is expected and related to the distances between Eu atoms (shorter in $ab$ plane than along $c$ direction).

To correlate the experimental data with the effective Heisenberg model, we have used two different spin blocks of size $8 \times 8 \times 16$ and $16 \times 16 \times 32$. 
The temperature dependence of the specific heat and average magnetization of a single layer is simulated with Monte Carlo method using {\sf SpinMC.jl} package~\cite{mccode}.
Results of our Monte Carlo simulations are presented in Fig.~\ref{fig:MC}.
The magnetic order parameter (defined as an average magnetization of the triangular Eu layer), falls abruptly around $21$~K [Fig.~\ref{fig:MC}(b)].
Around the same temperature, the specific heat also possesses a clear peak [Fig.~\ref{fig:MC}(c)], which directly indicates a magnetic phase transition from AFM to PM phase.
Theoretically obtained N\'eel temperature $T_{\rm N} \approx 21$~K is in  excellent agreement with experimental one.
Presented results weakly depend on the size of the investigated system [cf. red and blue lines on Fig.~\ref{fig:MC}(b, c)]. The presence of two exchange coupling parameters, with opposite sign, are also reflected in the magnetocaloric effect, which is shown in the SM~\cite{SuppMat}.

\paragraph*{Magnetic interaction mechanism.} 
We propose an extended superexchange mechanism~\cite{zhang.kong.19}, related to superposed electronic states of cation in a specific path, as a source of the magnetic interaction in EuZn$_{2}$P$_{2}$.
In the case of typical superexchange interaction, transition metal cations interact indirectly through intervening nonmetal anions.
The superexchange interaction, explained by Goodenough--Kanamori--Anderson rule~\cite{goodenough.63}, strongly depends on the cation-anion-cation bond angles. 
It was argued that the $180^{\circ}$ bond stabilizes an AFM ground state, while a $90^{\circ}$ bond is responsible for the FM interactions. 
Nevertheless, the mentioned mechanism is not limited by only said two bond angles, rather allowed to be extended to any bond angles. 
The extended version of the superexchange interaction was previously used to explain the magnetic ground state of several compounds, viz. CrO$X$ ($X=$ Cl, Br)~\cite{zhang.kong.19,woo.kiem.21}, or Rb$Ln$Se$_{2}$ ($Ln=$ Ce, Pr, Nd, Gd).~\cite{azzouz.halit.20}

In the case of EuZn$_{2}$P$_{2}$, the magnetic interaction is mediated by P atoms, which is allowed through two different paths [see Fig.~\ref{fig:dens}(a)].
In the $ab$ plane, the magnetic interaction is realized by ``path A'' (blue dashed line), containing one P atom.
Here, the Eu-P-Eu bond angle ($\approx 85.52^{\circ}$) is close to the ideal $90^{\circ}$ value, which favors the ferromagnetic coupling.
On the other hand, the magnetic interaction between two Eu atoms along the $c$-direction is mediated by two P atoms located between two Eu-layers and marked by ``path B'' (green dashed line), in Fig.~\ref{fig:dens}(a)
In this case, the Eu-P-P bond angle ($\approx 105.90^{\circ}$) 
is relatively large and corresponds to the AFM exchange coupling. Very similar bond lengths, (and hence the bond angles) of the Zintl phases simultaneously possessing insulating, and AFM properties are discussed in Ref.~\cite{berry.stewart.22}, although without any exchange interactions. The authors of that paper phenomenologically approximated the strength ratio of the AFM and FM interactions as  1:-4. We accounted for the Heisenberg exchange interactions $J_{\perp}$ and $J_{\parallel}$, and showed that $J_{\parallel}/J_{\perp}$ is very close to -3.5. Thus, we propose that the extended superexchange mechanism can be used to explain the AFM ordering in insulating Zintl phases, that possess strong intralayer FM and weak interlayer AFM couplings. 
Moreover, such extended superexchange mechanism should be predominant in this class of materials.

To further strengthen our argument, we have calculated the modification of charge as well as spin density distribution of the crystal using the DFT technique.
Here, we can define modification of the charge density as~\cite{ptok.kapcia.21}:
\begin{eqnarray}
\Delta \rho = \rho_\text{EuZn$_{2}$P$_{2}$} - \left( \rho_\text{Eu} + \rho_\text{Zn$_{2}$} + \rho_\text{P$_{2}$} \right) ,
\end{eqnarray}
where $\rho_\text{EuZn$_{2}$P$_{2}$}$ denotes density distribution for EuZn$_{2}$P$_{2}$, while $\rho_{i}$ density distribution for separated component forming this compound (i.e., $i =$ Eu, Zn, and P).
The spin density redistribution can be defined in an analogous way.
Fig.~\ref{fig:dens}(b) shows a transfer of charge density mostly between Zn and P, which is related to the realization of strong bonding between these atoms.
Nevertheless, the electron density around Eu and Zn decreases and increases around P. 
This suggests the anion-like character of the phosphorus in chemical bonding of EuZn$_{2}$P$_{2}$, which we predicted in the previous paragraph applying of extended superexchange scheme.
Interestingly, a contrasting behavior is observed in the case of the spin density rearrangement [see Fig.~\ref{fig:dens}(c)].
In this case, we observed a transfer of the spin density only between Eu and P atoms.
Indeed, DFT analysis indicates the occurrence of the small magnetic moment on P atoms (with magnitude $\approx 0.02$~$\mu_{\rm B}$).
Observed spin density redistribution between Eu and P atoms, suggests that the P atoms play the role of the mediator within the magnetic interaction mechanism. 
Such rearrangement of the spin density further supports our choice of virtual hopping path (marked with red and green in Fig.~\ref{fig:dens}(a)).
Additionally, we notice, that the electronic states closest to the ``Fermi level'' (top of valence band) are realized mostly by the Zn and P orbitals, what is well visible on the electronic density of states projected on atoms (see Fig.~S5 in SM~\cite{SuppMat}). This finding is in agreement with earlier investigation of EuZn$_{2}$P$_{2}$~\cite{berry.stewart.22}.
This further supports out prediction about the role of Zn and P states in the mechanism of magnetic interaction, and forming of the AFM order.

\begin{figure}
    \centering
    \includegraphics[width=\columnwidth]{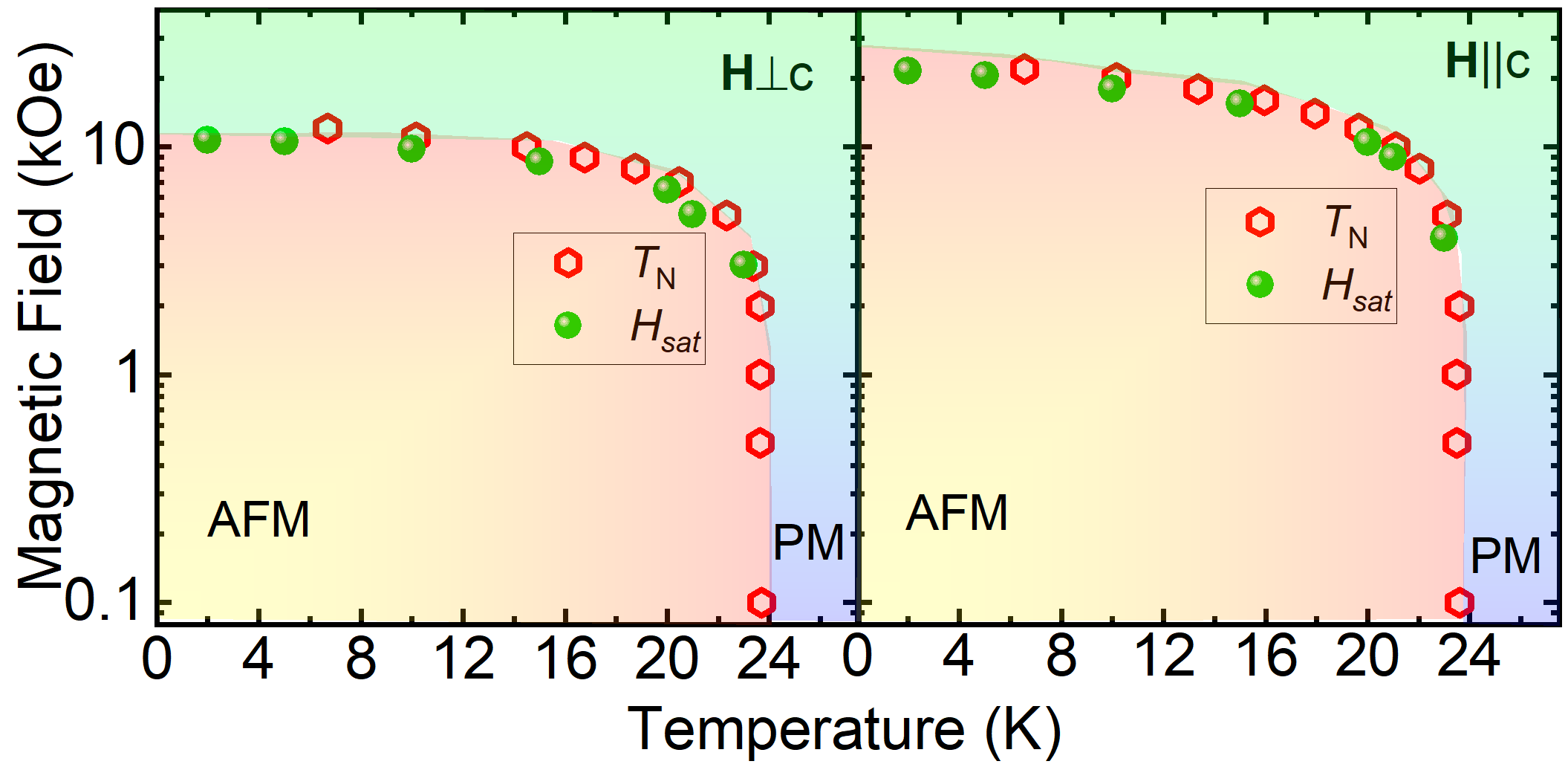}
    \caption{Magnetic phase diagram of EuZn$_{2}$P$_{2}$, drawn for magnetic field applied perpendicular and parallel to the $c$-axis, using values of N\'eel temperature (red points) and saturation field (green points).}
   \label{fig:phase}
\end{figure}

\section{Summary}
\label{sec.sum}

The temperature and magnetic field-dependent magnetic susceptibility, isothermal magnetization, magnetocaloric effect, heat capacity, and electrical resistivity of EuZn$_{2}$P$_{2}$ single crystal indicate an antiferromagnetic phase transition at $T_{\rm N} = 23.5$~K due to the ordering of Eu$^{2+}$ spins. 
$M(H)$ at $T=2$~K, saturates to  $6.8$~$\mu_{\rm B}$ at $H_{sat}^{\perp} = 14$~kOe for \textbf{H}$\perp$\textit{c} and at the $H_{sat}^{\parallel} = 24$~kOe for \textbf{H}$\parallel$\textit{c} directions. 
The magnetic entropy retrieved from the heat capacity data is equal to $17$~J mol$^{-1}$K$^{-1}$, which closely matches the theoretical one. 

$M(H)$ confirms that the easy-axis lies in the $ab$-plane. 
Electrical resistivity reveals the insulating nature of EuZn$_{2}$P$_{2}$, with band gap $\approx 0.17$\,eV, which survives (only slightly reduced), up to $140$~kOe. 
Based on the magnetization data in function of both temperature and magnetic field, we sketched a magnetic phase diagram of the compound for both \textbf{H} $\perp$\textit{c} and \textbf{H} $\parallel$\textit{c} directions. (Fig.~\ref{fig:phase}). $T_{\rm N}$ and $H_{sat}$ were obtained from $\chi(T)$ and $M(H)$ data, respectively.

We proposed an extended superexchange interaction to be responsible for the magnetic ordering. 
The charge and spin density modifications indicate that phosphorus atoms play the role of anionic mediators, necessary in the superexchange mechanism.
We formulated an effective model of the magnetic interaction, based on the {\it ab initio} DFT calculations.
We found that the interaction in $ab$ plane, realized via the Eu-P-Eu path, with bond angle of $\approx86^{\circ}$ is moderately ferromagnetic ($J_{\parallel} = 1.23$~meV), while that along $c$ direction (Eu-P-P-Eu path, with bond angle $\approx106^{\circ}$) is weakly antiferromagnetic ($J_{\perp} = -0.35$~meV).
The experimentally determined N\'eel temperature was successfully reproduced by the Monte Carlo simulations with those interaction parameters, which strongly supports the superexchange mechanism we propose.
Theoretical calculation of the magneto-crystalline anisotropy energy of the compound in the $ac$ plane reveals that the easy axis is along the crystallographic $a$ direction.
This finding is in agreement with a recent investigation~\cite{krebber.kopp.23}.
Our results shed also new light on the mechanism of magnetic interactions in the Eu based compounds, like EuZn$_{2}$P$_{2}$.

\begin{acknowledgments}
Some figures in this work were rendered using {\sc Vesta}~\cite{momma.izumi.11}.
We would like to thank F. L. Buessen for sharing his Monte Carlo code. 
This work was supported by National Science Centre (NCN, Poland) under Projects No. 2021/41/B/ST3/01141 (K.S., P.W., and D.K.) and 2021/43/B/ST3/02166 (A.P.). 
A.P. is grateful to Laboratoire de Physique des Solides in Orsay (CNRS, University Paris Saclay) for hospitality during a part of the
work on this project.
\end{acknowledgments}

\bibliography{biblio.bib}


\clearpage
\newpage

\onecolumngrid

\begin{center}
  \textbf{\Large Supplemental Material}\\[.2cm]
  \textbf{\large Superexchange Interaction in Insulating EuZn$_{2}$P$_{2}$}\\[.2cm]
 Karan Singh$^{1}$, Shovan Dan$^{1}$, A. Ptok$^{2}$, T. A. Zaleski$^{1}$, O. Pavlosiuk$^{1}$, P. Wi\'{s}niewski$^{1}$, and D. Kaczorowski$^{1}$\\[.2cm]
  {\itshape
  	\mbox{$^{1}$Institute of Low Temperature and Structure Research, Polish Academy of Sciences, Okólna 2, 50-422 Wrocław, Poland}\\
	\mbox{$^{2}$Institute of Nuclear Physics, Polish Academy of Sciences, W. E. Radzikowskiego 152, PL-31342 Kraków, Poland}
  }
(Dated: \today)
\\[1cm]
\end{center}

\setcounter{equation}{0}
\renewcommand{\theequation}{S\arabic{equation}}
\setcounter{figure}{0}
\renewcommand{\thefigure}{S\arabic{figure}}
\setcounter{section}{0}
\renewcommand{\thesection}{S\arabic{section}}
\setcounter{table}{0}
\renewcommand{\thetable}{S\arabic{table}}
\setcounter{page}{1}

\begin{figure}[h]
    \centering
    \includegraphics[width=0.6\textwidth]{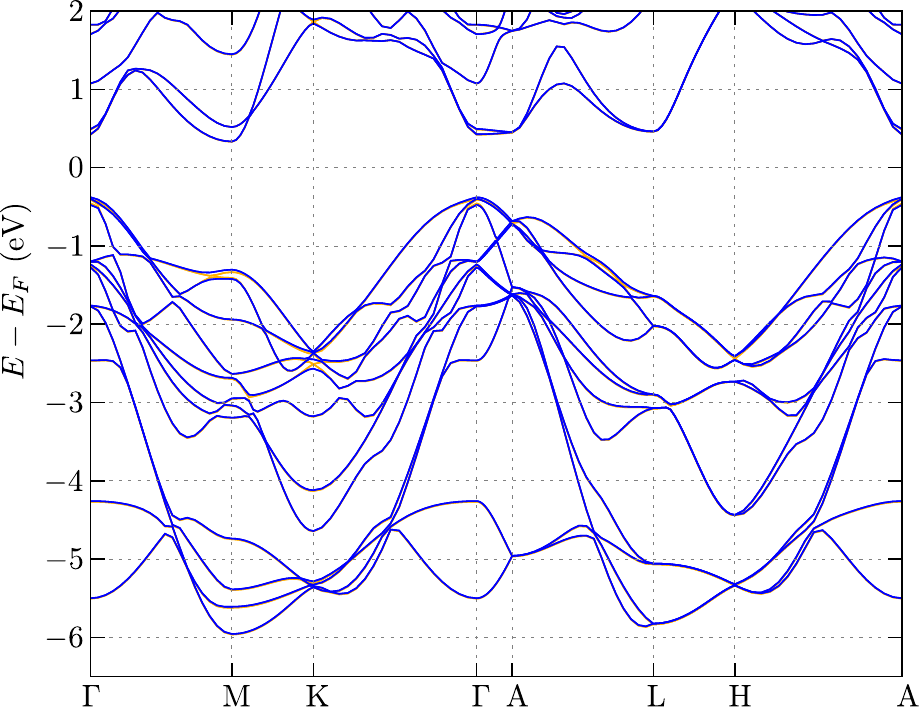}
    \caption{The electronic band structure of  EuZn$_{2}$P$_{2}$ with Eu $4f$ electrons treated as core states.
    The orange and blue lines present results in the absence and presence of the spin--orbit coupling, respectively.}
    \label{fig:band_nof}
\end{figure}
\begin{figure}
    \centering
    \includegraphics[width=0.525\textwidth]{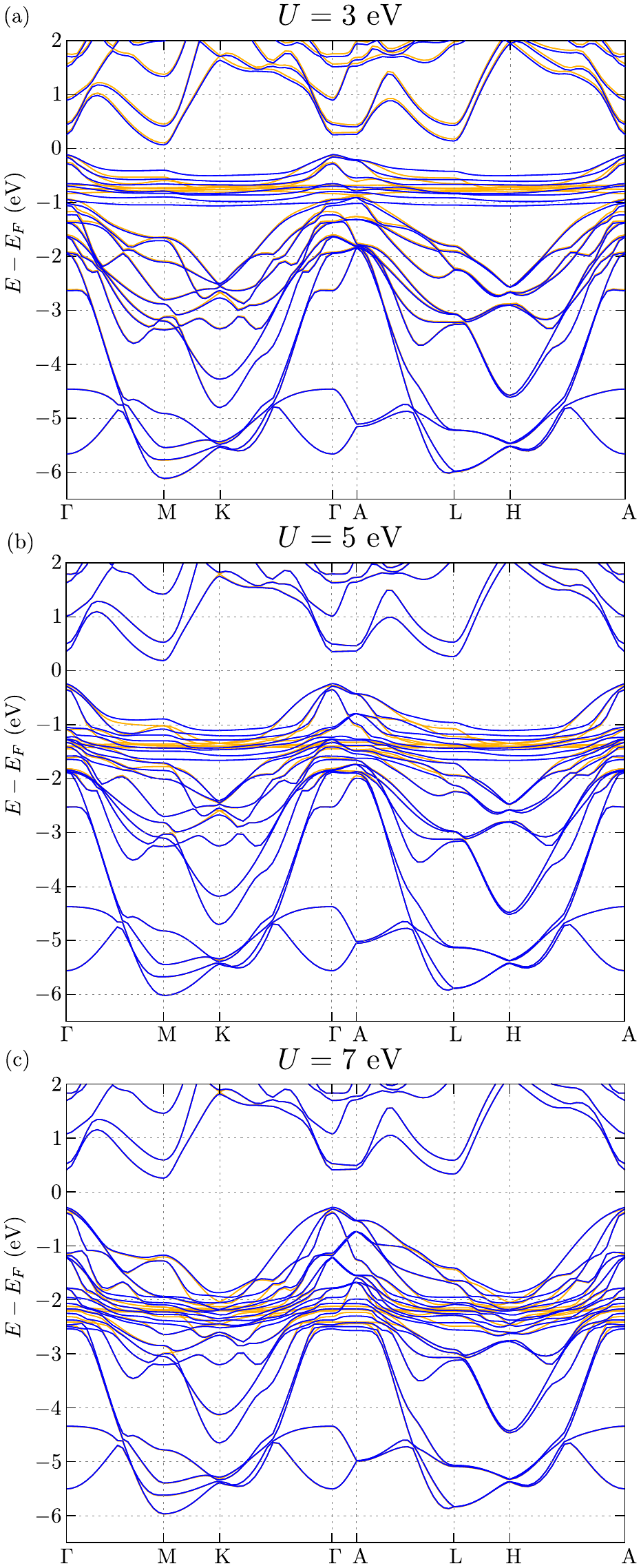}
    \caption{Influence of the Hubbard $U$ on the electronic band structure.
    Results for $U$ equal $3$\,eV, $5$\,eV, and $7$\,eV (from top to bottom).
    The orange and blue lines present results in the absence and presence of the spin--orbit coupling, respectively.}
    \label{fig:band_u}
\end{figure}
\begin{figure}
    \centering
    \includegraphics[width=0.6\columnwidth]{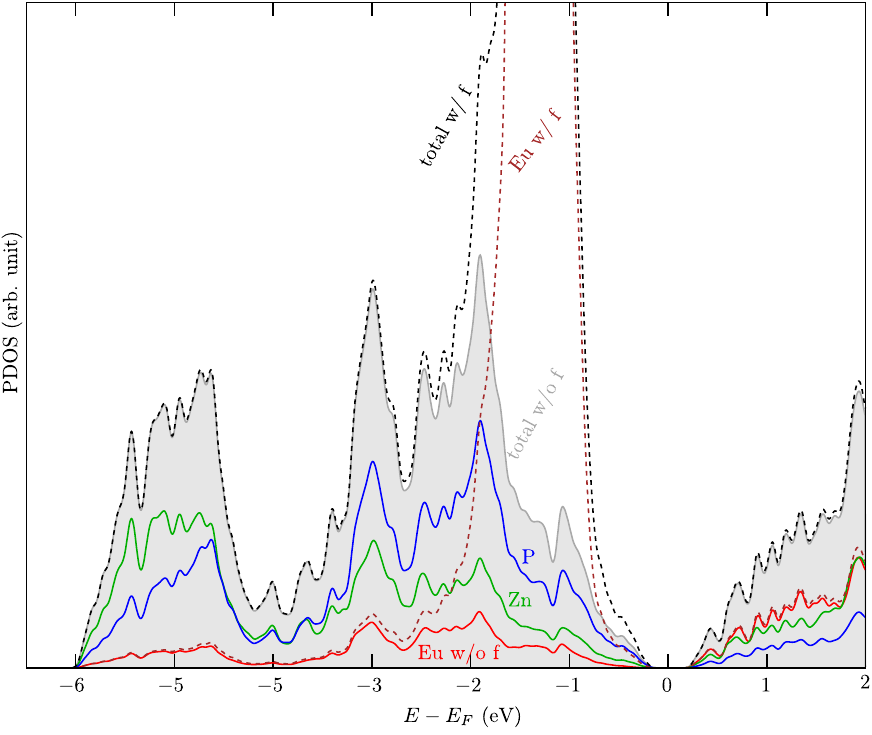}
    \caption{
    Electronic partial density of states (PDOS). 
    The shaded gray area corresponds to the total DOS for the $f$ orbitals treated as core states.
    Red, green, and blue color solid lines correspond to the PDOS of Eu, Zn, and P, respectively.
    Similarly, dashed black and brown lines, correspond to the total DOS and Eu PDOS, respectively, when $f$ orbitals are treated as valence states.}
    \label{fig:eldos}
\end{figure}
\begin{figure}[h]
    \centering
    \includegraphics[width=0.7\columnwidth]{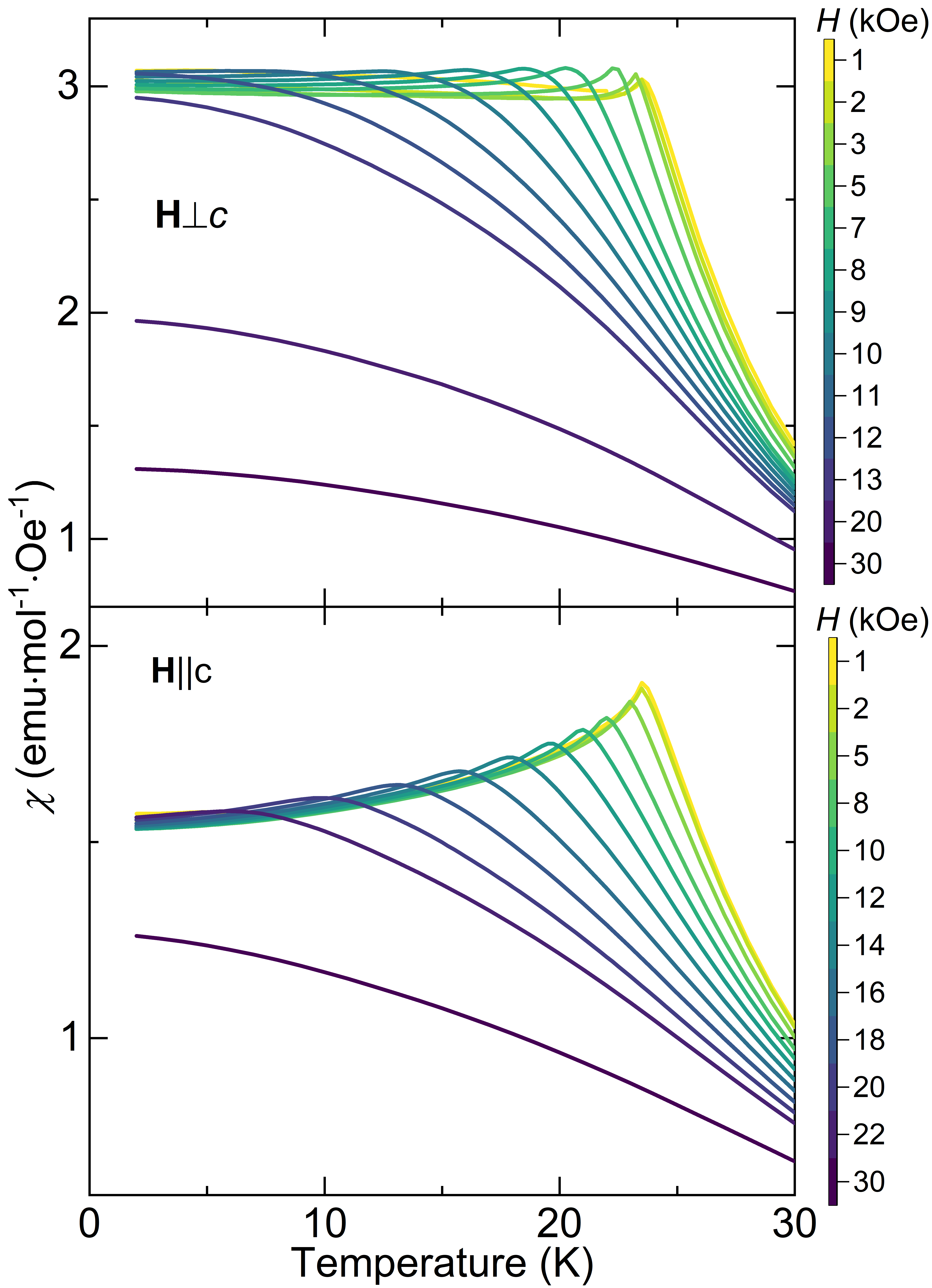}
    \caption{Magnetic susceptibility measured in different magnetic fields applied (a) perpendicular and (b) parallel to the $c$-axis.}
    \label{fig:chi}
\end{figure}
\begin{figure}[h]
    \centering
    \includegraphics[width=0.7\columnwidth]{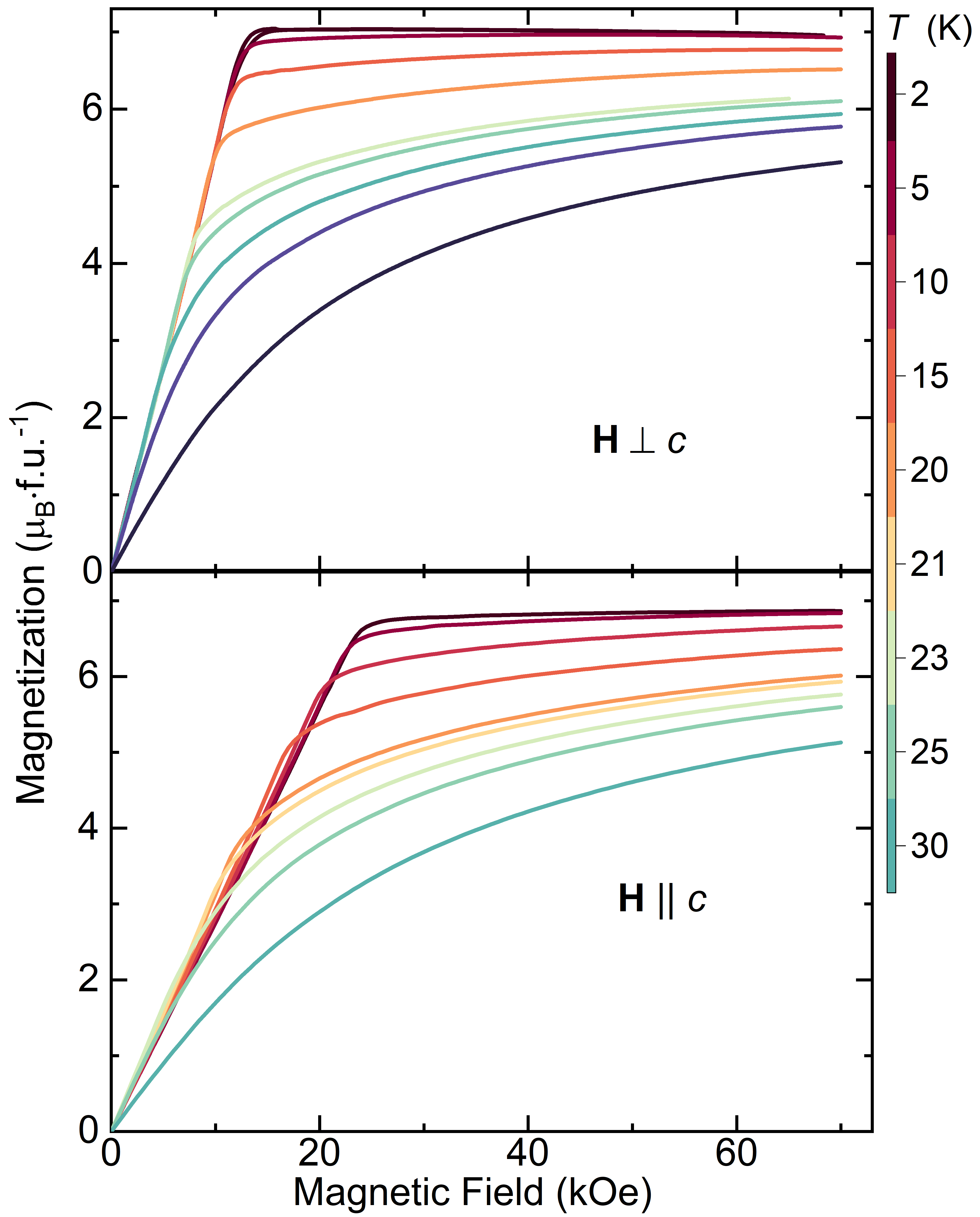}
    \caption{Magnetic isotherms at different temperatures in magnetic fields applied (a) perpendicular and (b) parallel, to the $c$-axis.}
    \label{fig:MH}
\end{figure}

In this Supporting Information, we present additional results:
\begin{itemize}
\item Sec.~\ref{sm.structure} Structure and composition characterization.
\item ~~~~~~Fig.~\ref{fig:laue} Laue diffractogram of the EuZn$_{2}$P$_{2}$.
\item ~~~~~~Fig.~\ref{fig:eds} EDS spectrum.
\item  Sec.~\ref{sm.dft_mc} Exchange parameters based on DFT calculations.
\item ~~~~~~Fig.~\ref{fig:band_nof} The electronic band structure of EuZn$_{2}$P$_{2}$ with Eu $4f$ electrons treated as core states.
\item ~~~~~~Tab.~\ref{tab.gaps} Band gaps from DFT calculations dependently by $U$ value.
\item ~~~~~~Fig.~\ref{fig:band_u} Influence of the Hubbard $U$ on the electronic band structure ($f$ electrons treated as valence states).
\item ~~~~~~Fig.~\ref{fig:eldos} Electronic partial density of states (PDOS).
\item Sec.~\ref{sm:m} Magnetic susceptibility and magnetization.
\item ~~~~~~Fig.~\ref{fig:chi} Magnetic susceptibility.
\item ~~~~~~Fig.~\ref{fig:MH} Magnetic isotherms at different temperatures.
\item Sec.~\ref{sm:cp} Specific heat.
\item ~~~~~~Fig.~\ref{fig:cp_decomposed} $C_{P}/T^3$  and its fit using Debye and Einstein contribution.
\item ~~~~~~Fig.~\ref{fig:cp_reply} $C_{P}/T$ fitted using two Debye model and concerned $S_{mag}$.
\item Sec.~\ref{ssec:MCE}  Magnetocaloric effect.
\item ~~~~~~Fig.~\ref{fig:mce} $\Delta S_M(T)$  in \textbf{H}$\parallel c$, and \textbf{H}$\perp c$ directions.
\end{itemize}

\newpage

\section{Structure and composition characterization}
\label{sm.structure}

\begin{figure}[h]
    \centering
    \includegraphics[width=0.6\textwidth]{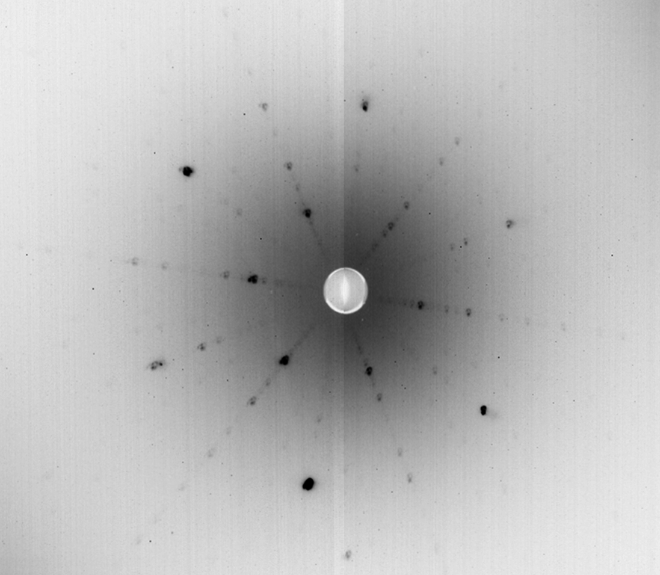}
    \caption{Laue diffractogram of the EuZn$_{2}$P$_{2}$ crystal obtained with incident X-ray beam along [001] direction.}
    \label{fig:laue}
\end{figure}

\begin{figure}[h!]
    \centering
    \includegraphics[width=0.6\textwidth]{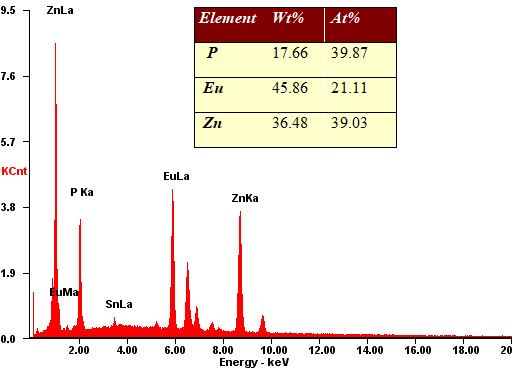}
    \caption{Energy-dispersive X-ray spectrum collected on arbitrary spot on the EuZn$_{2}$P$_{2}$ crystal. Weight and atomic composition are shown in the inset.}
    \label{fig:eds}
\end{figure}

To analyze the chemical composition, Energy-dispersive X-ray spectroscopy (EDS) experiment was carried out on a few pieces of the synthesized EuZn$_{2}$P$_{2}$ crystal, which shows that stoichiometry is a good agreement with the nominal composition (see Fig.~\ref{fig:eds}). The Laue diffractogram of the oriented EuZn$_{2}$P$_{2}$ crystal is shown in Fig.~\ref{fig:laue}, it perfectly corresponds to the trigonal structure with space-group $P\bar{3}m1$.

\section{Exchange parameters based on DFT calculations}
\label{sm.dft_mc}

To find the parameters of exchange  between magnetic moments of Eu, we performed DFT calculation within the supercell containing $2\times 2\times 1$ magnetic unit cells.
We calculate the ground states (GS) energies for four fixed configurations of magnetic moments:
\begin{itemize}
\item AFM configuration, with the GS energy $-211.38165$\,eV,
\item AFM configuration with one inverted magnetic moment, and the GS energy $-211.36549$\,eV,
\item FM configuration, with the GS energy $-211.37605$\,eV,
\item FM configuration with one inverted magnetic moment, and the GS energy $-211.36269$\,eV.
\end{itemize}
In the simplest case, the GS energy of the mentioned configurations may be used to represent the total energy of the system as a function of the magnetic moments, in the form:
\begin{eqnarray}
E = E_{0} - \sum_{ij} J_{\parallel} {\bm e}_{i} {\bm e}_{j} - \sum_{ij} J_{\perp} {\bm e}_{i} {\bm e}_{j} ,
\end{eqnarray}
where summation over sites $i$ and $j$ is perform within the FM layer ($J_{\parallel}$ term) or between AFM layers ($J_{\perp}$ term).
$E_{0}$ denotes the energy of the system in the absence of magnetic order (i.e. non-magnetic system).
Solving the system of equations for the mentioned configuration, we found $E_{0} = -211.349$\,eV, $J_{\parallel} = 1.23$~meV, and $J_{\perp} = -0.35$~meV.

\begin{table*}[h]
\caption{
\label{tab.gaps}
Band gaps from DFT calculations
}
\begin{tabular}{ccc}
Hubbard $U$ & indirect gap (eV) & direct gap (eV) \\
\hline
w/o U ($f$ states in core) -- Fig.~\ref{fig:band_nof} & $0.7120$ & $0.8047$ \\
$U = 3$~eV -- Fig.~\ref{fig:band_u}(a) & $0.1784$ & $0.3678$ \\
$U = 5$~eV -- Fig.~\ref{fig:band_u}(b) & $0.4251$ & $0.5933$ \\
$U = 7$~eV -- Fig.~\ref{fig:band_u}(c) & $0.5412$ & $0.6953$ \\
\end{tabular}
\end{table*}

\clearpage
\newpage

\section{Magnetic susceptibility and magnetization}\label{sm:m}

\clearpage
\newpage

\section{Specific heat}\label{sm:cp}
\begin{figure}[!h]
\centering
\includegraphics[width=0.55\textwidth]{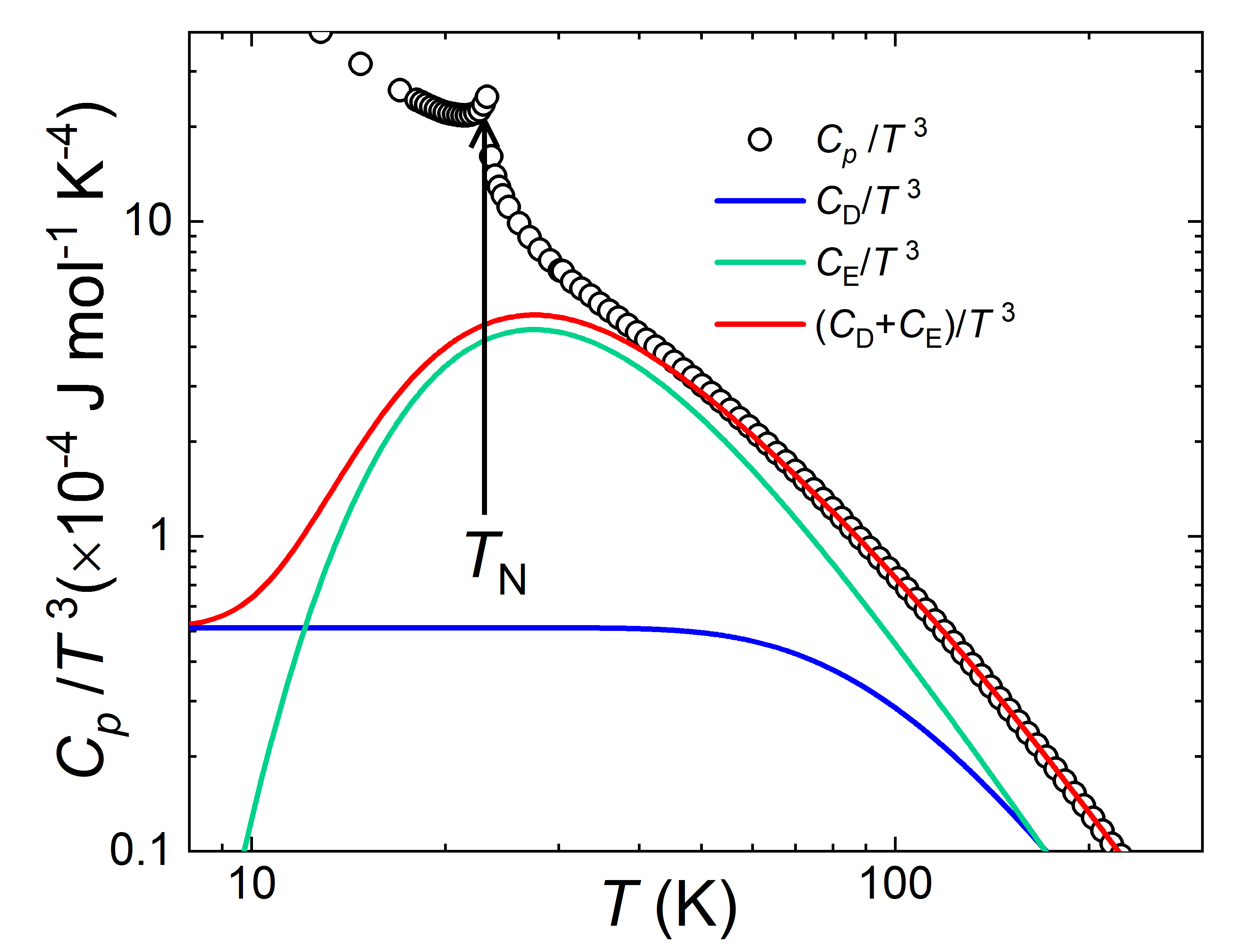}
\caption{$C_{P}/T^3$  and its fit using Debye and Einstein contribution. Decomposed Debye and Einstein contribution ($C_{\rm D}/T^3$ and $C_{\rm E}/T^3$) are also plotted as a function of $T$.
    }
\label{fig:cp_decomposed}
\end{figure}

For a compound containing Einstein contribution, it is expected to have a hump in the $C_{p}/T^3$ vs. $T$ plot. However, in the present case, the hump occurs close to the magnetic ordering temperature and not properly visible in naked eye. For visualizing the feature, we have decomposed the specific heat data into the Debye and Einstein, as shown in Fig. \ref{fig:cp_decomposed} and plotted the $C_{\rm p}/T^3$ vs. $T$ plot, which clearly shows the hump near to the ordering temperature due to Einstein contribution.

\begin{figure}[h]
\centering
\includegraphics[width=0.55\textwidth]{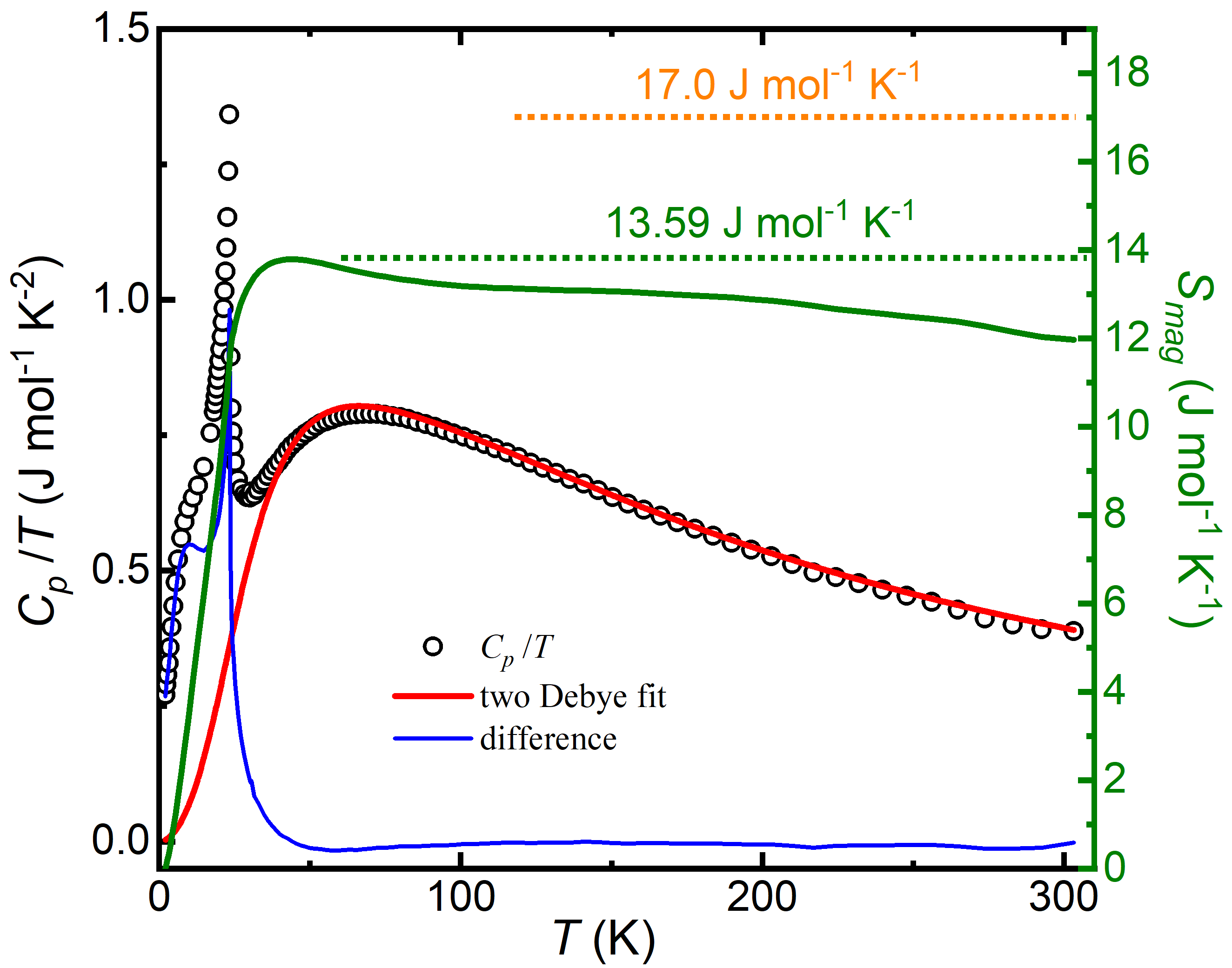}
\caption{Left-axis: $C_p/T$ fitted to (in the range 40--300~K) the two-Debye model as proposed in Ref.~12. Fitting residuals are plotted with a blue line. Right-axis: the magnetic entropy. The entropy values obtained with the two-Debye-terms model and our with fit including one Debye and one Einstein mode are shown with olive and orange dotted line, respectively. The actual magnetic entropy for Eu$^{2+}$ is $\mathcal{R}\ln8 = $17.3 J mol$^{-1}$ K$^{-1}$.
    }
\label{fig:cp_reply}
\end{figure}

We have also shown that the two-Debye model, as  proposed in ref: 12, does not able to rescue the full magnetic entropy of Eu$^{2+}$ ions (Fig. \ref{fig:cp_reply})

\clearpage
\newpage

\section{Magnetocaloric effect}
\label{ssec:MCE}

The magnetocaloric effect (MCE) is an important phenomenon providing an overview of underlying magnetic interactions, apart from its importance for technological applications. For the latter aspect, the absolute value of magnetic entropy change ($|\Delta S_M|$), and the relative cooling power play crucial roles. However, recently, MCE is being also used as a tool to explain the complex competition of magnetic interactions, eg. in skyrmions~\cite{jamaluddin2019robust}.
The magnetic entropy gain upon change of magnetic field from $H_1$ to $H_2$, can be obtained from Maxwell's relation:
\begin{equation}
    \Delta S_M (T, \Delta H) = \int_{H_1}^{H_2}\left(\frac{\partial M}{\partial T}\right ) {\rm d}H.
    \label{eqn:mce}
\end{equation}
Nevertheless, the magnetic entropy change is positive for FM interactions and negative for AFM interactions, dominating in given field range, and temperature ~\cite{jamaluddin2019robust}.

\begin{figure}[h]
    \centering
    \includegraphics[width=0.5\columnwidth]{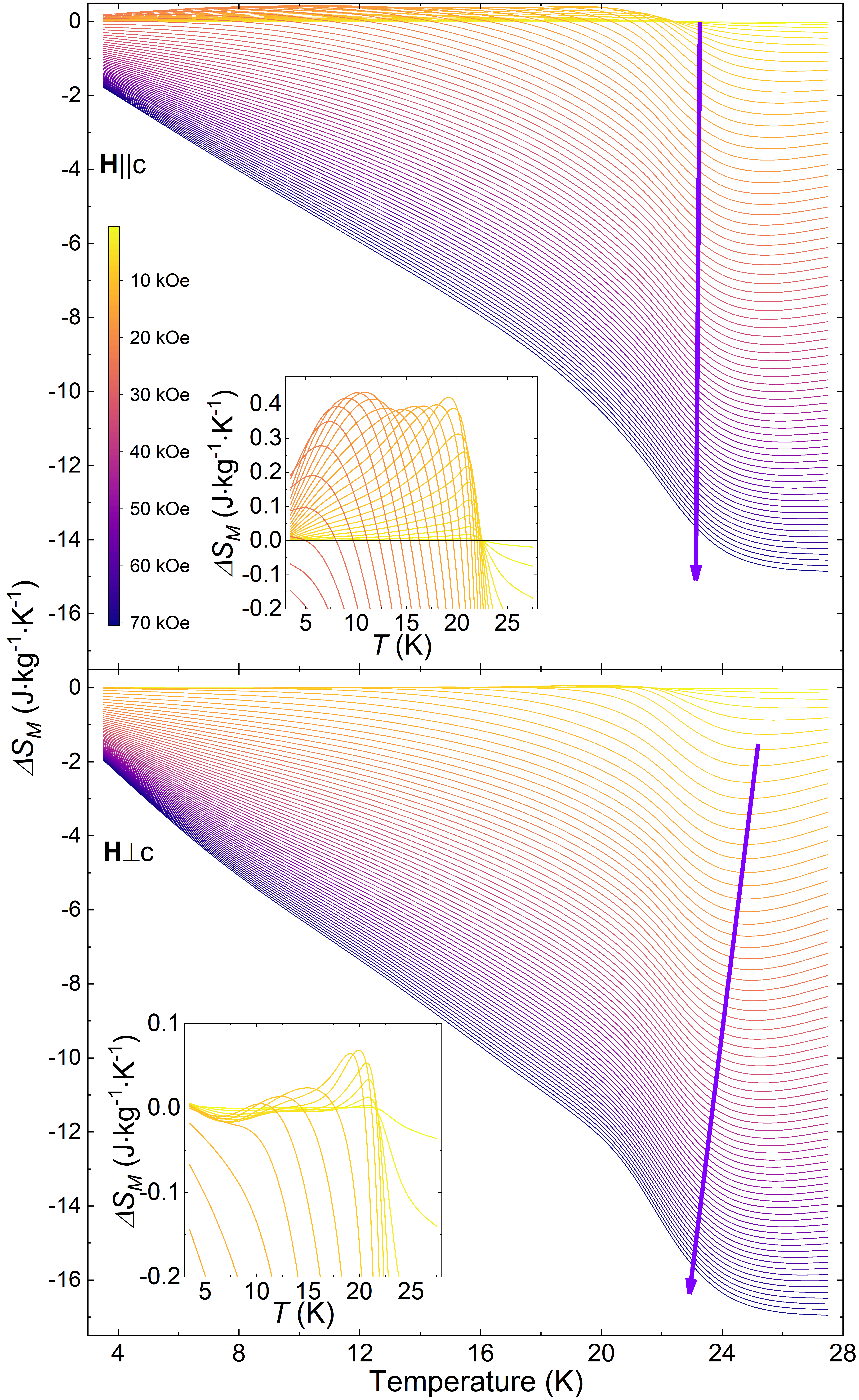}
    \caption{$\Delta S_M(T)$ calculated in magnetic fields (a) perpendicular and (b) parallel to the $c$-axis. Insets show the field region $0 \rightarrow 25$ kOe for \textbf{H}$\parallel$\textit{c} and $0 \rightarrow 10$ kOe for \textbf{H}$\perp$\textit{c} directions.}
    \label{fig:mce}
\end{figure}

Fig.~\ref{fig:mce} shows the $\Delta S_M$ for both \textbf{H}$\perp$\textit{c} and \textbf{H}$\parallel$\textit{c} field directions. $|\Delta S_M|$ attains maxima near $T_{\rm N}$ with $\approx$ 14.9 and 17 J$\cdot$kg$^{-1}$K$^{-1}$, for magnetic field applied in perpendicular and parallel to $c$-axis, respectively. 
Closer observation of $\Delta S_M$ at the low fields (see the insets of Fig.~\ref{fig:mce}), indicates a positive contribution in the $\Delta S_M$ value, below $T_{\rm N}$, and thus the domination of FM interactions, in full agreement  with our discussion of magnetization.

\end{document}